\documentclass[superscriptaddress,reprint,amsmath,amssymb,aps,prl,footinbib]{revtex4-1}
\usepackage{graphicx}
\usepackage{bm}
\usepackage{color}
\usepackage{hyperref}
\usepackage{caption}
\hypersetup{colorlinks,citecolor=blue}
\newcommand{\ket}[1]{\mbox{$|#1\rangle$}}
\newcommand{\bra}[1]{\mbox{$\langle #1|$}}

\newcommand{\+}{$^+$}
\newcommand{\TeH}{TeH$^+$}

\newcommand{\tj}[6]{ \begin{pmatrix}
   #1 & #2 & #3 \\
   #4 & #5 & #6
\end{pmatrix}}

\begin{document}

\title{Prospects for Polar Molecular Ion Optical Probe of Varying Proton-Electron Mass Ratio}

\author{Mark G. Kokish}
\affiliation{Department of Physics and Astronomy, Northwestern University, Evanston, Illinois 60208, USA}
\author{Patrick R. Stollenwerk}
\affiliation{Department of Physics and Astronomy, Northwestern University, Evanston, Illinois 60208, USA}
\author{Masatoshi Kajita}
\affiliation{National Institute of Information and Communications Technology, Koganei, Tokyo 184-8795, Japan}
\author{Brian C. Odom}
\email{b-odom@northwestern.edu}
\affiliation{Department of Physics and Astronomy, Northwestern University, Evanston, Illinois 60208, USA}

\date{\today}
\begin{abstract}
Molecules with deep vibrational potential wells provide optical intervals sensitive to variation in the proton-electron mass ratio ($\mu$).  On one hand, polar molecules are of interest since optical state preparation techniques have been demonstrated for such species.  On the other hand, it might be assumed that polar species are unfavorable candidates, because typical molecule-frame dipole moments reduce vibrational state lifetimes and cause large polarizabilities and associated Stark shifts.  Here, we consider single-photon spectroscopy on a vibrational overtone transition of the polar species TeH$^+$, which is of practical interest because its diagonal Franck-Condon factors should allow rapid state preparation by optical pumping.  We point out that all but the ground rotational state obtains a vanishing low-frequency scalar polarizability from coupling with adjacent rotational states, because of a fortuitous relationship between rigid rotor spacings and dipole matrix elements.
We project that for good choices of spectroscopy states, demonstrated levels of field control should make possible uncertainties of order $1 \times 10^{-18}$, similar to those of leading atomic ion clocks.  The moderately long lived vibrational states of TeH$^+$ make possible a frequency uncertainty approaching $1 \times 10^{-17}$ with one day of averaging for a single trapped ion.  Observation over one year could probe for variation of $\mu$ with a sensitivity approaching the $1 \times 10^{-18}/\textrm{yr}$ level.
\end{abstract}
\maketitle

Searches for variation of fundamental constants are motivated by their ability to probe physics beyond the Standard Model~\cite{Uzan2011}. Modern laboratory searches for variation use precise frequency measurements with sensitivity to the fine structure constant ($\alpha$) and the proton-electron mass ratio ($\mu$)~\cite{Jansen2014}. Improved searches for variation of $\mu$ are especially intriguing as it is predicted to drift faster than $\alpha$ in generic models~\cite{Flambaum2004}. If astronomical observations of methanol lines are cast in terms of a linear temporal drift in $\mu$, they set a limit of $2.4 \times 10^{-17}$/yr~\cite{Bagdonaite2013}. The tightest laboratory constraint on the fractional variation of $\mu$, $\sim$$1 \times 10^{-16}$/yr, was obtained from a comparison of hyperfine and electronic transitions in atomic clocks~\cite{Godun2014,Huntemann2014}, using a shell model calculation to describe the dependence of the nuclear magnetic moment on $\mu$~\cite{Flambaum2006}. Since the sensitivity to $\mu$ arises from the relatively low frequency ($\sim 10$ GHz) hyperfine transition, it will be challenging to significantly improve the precision of $\mu$ variation searches by this approach. Vibrational transitions in molecules provide model-independent sensitivity to varying $\mu$, with the current best constraint ($< 5.6 \times 10^{-14}$/yr) obtained in a molecular beam~\cite{Shelkovnikov2008}.

Spectroscopy on single trapped atomic ions has achieved statistical and systematic uncertainties at the low $10^{-18}$ level~\cite{Huntemann2016, chen_sympathetic_2017}.  Recent demonstrations of molecular ion quantum state preparation~\cite{Staanum2010, Schneider2010, Hansen2014, Lien2014, Chou2017} and non-destructive readout~\cite{Chou2017,Wolf2016} suggest that spectroscopy on a single trapped molecular ion could obtain a high duty cycle in an environment also favorable for control of systematic uncertainties.  In order to evaluate whether this approach to molecular spectroscopy could improve $\mu$ variation sensitivity beyond the $10^{-16}$ level of atoms, the intrinsic details of the molecular states and practical aspects of state preparation must be carefully considered.

Compared with hyperfine transitions in atoms, high vibrational overtone intervals (10-1000 THz) of molecules have orders of magnitude larger absolute sensitivity to varying $\mu$~\cite{Demille2008,Hanneke2016}.
Optical-frequency single-photon overtone transitions have been observed in trapped molecular ions ~\cite{Koelemeij2012, Khanyile2015}. When the state lifetimes are sufficiently long, such overtone transitions offer a means to surpass the statistical sensitivity of previous searches.  One proposed technique is to drive a transition from a vibrationally excited state to a nearly degenerate level with different $\mu$ sensitivity~\cite{Demille2008, Kozlov2009, Chin2006, Flambaum2007, Hanneke2016}, thereby benefiting from relaxed requirements for the probe stability and a suppressed absolute Doppler shift~\cite{Demille2008, Hanneke2016}.
A challenge of this approach is to find suitable transitions where the dissimilar character of the states does not cause large differential shifts and systematic uncertainties. An alternative approach is to measure the vibrational overtone frequency directly by one-photon~\cite{schiller_simplest_2014, karr_candidates_2014} or two-photon~\cite{Zelevinsky2008, Kajita2012,kajita_test_2014,karr_candidates_2014,kajita_search_2017} transitions.

Systematic frequency shifts in polar molecules will generally arise from coupling of nearby rotational and vibrational levels, a serious concern absent in atoms. One response is to use nonpolar (homonuclear) diatomic molecules, whose vanishing dipole moment eliminates Stark shifts from rotational and vibrational coupling~\cite{kajita_test_2014, schiller_simplest_2014, karr_candidates_2014}.  However, it is also of great interest to consider polar molecules, particularly since demonstrated optical pumping state preparation techniques require a dipole moment~\cite{Staanum2010,Schneider2010} or a structure not yet identified in a homonuclear species~\cite{Lien2014}.  Polar molecules have closely spaced levels of opposite parity, which for example allow for molecular orientation in moderate electric fields.  One might naively expect that the associated Stark shifts would pose possibly catastrophic challenges for clock-level spectroscopy on polar species.  It has previously been pointed out for HD$^+$ that the DC scalar polarizability of rotationally excited states is in fact dominated by electronic couplings~\cite{schiller_static_2014}.  Other systematic uncertainties were considered in detail~\cite{schiller_simplest_2014, karr_candidates_2014}, and it was proposed that a weighted average over a carefully chosen set of disparate transitions could create a composite frequency with a low inaccuracy~\cite{ schiller_simplest_2014}.  Here, we point out that the remarkable feature of small DC scalar polarizability actually arises from a nearly precise cancellation of adjacent-level interactions, and that there is a related cancellation of the differential polarizability in the high frequency limit.  Recognizing that the only large polarizabilities unavoidable in polar molecules are tensorial in character, it becomes clear that simple state averaging techniques, known from atomic clocks, can be used to simultaneously suppress this shift as well as linear Zeeman and quadrupole shifts.

We consider the prospects of spectroscopy on a single TeH$^+$ ion for an improved search for varying $\mu$.  Several favorable properties of TeH$^+$ stem from its electronic structure, which has been recently calculated~\cite{GoncalvesdosSantos2015}, but has not yet been experimentally measured. The lowest few electronic states of TeH$^+$, X$_10^+$, X$_21$, a2, and b$0^+$, have diagonal Franck-Condon Factors (FCFs). These diagonal FCFs arise because the states correspond to different orbital and spin configurations of two electrons in non-bonding p orbitals localized on the tellurium ion, and inter-transitions leave the bond length and strength relatively unperturbed.  A diagonal transition from the ground state makes possible rapid spectroscopy state preparation by optical pumping~\cite{Lien2014, Stollenwerk2015}. Furthermore, diagonal FCFs reduce shifts arising from the upper spectroscopy state coupling to levels in other electronic manifolds which are close in energy but have poor vibrational overlap.

\begin{figure}
\includegraphics[scale=1.0]{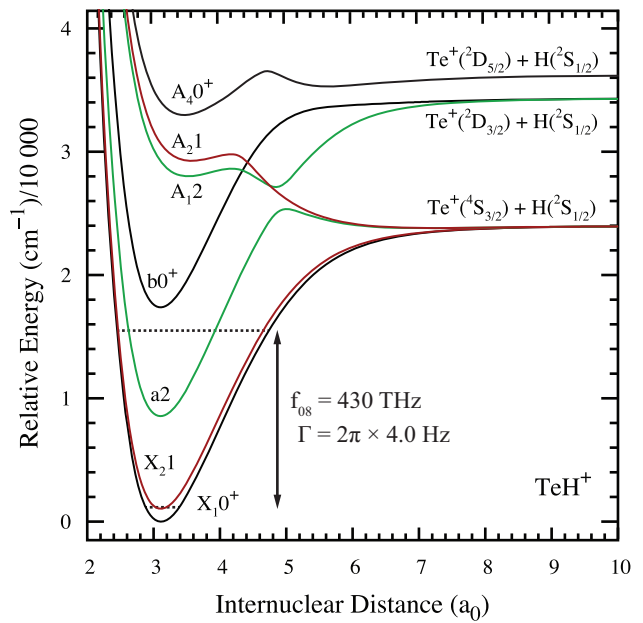}
\caption{\footnotesize Relevant low-lying electronic states of TeH$^+$. Dashed lines indicate the initial and final states of the spectroscopy transition, $v=0 \rightarrow v'=8$ in the X$_10^+$ manifold. Figure adapted from~\cite{GoncalvesdosSantos2015}.}
\label{fig:pec}
\end{figure}

\TeH\ is polar with a ground-state body-frame dipole moment of 0.91 Debye.  In the absence of spin-orbit coupling, the TeH$^+$ ground state is $^3\Sigma ^-$, and the two lowest excited states correspond to $^1\Delta$ and $^1\Sigma^+$ states.  However, strong spin-orbit coupling originating from the heavy tellurium atom makes the Hund's case (c) basis a good approximation to the eigenstates~\cite{Brown2003} (Fig.~\ref{fig:pec}).  Selection rules that would otherwise prevent transitions between the three low lying states in the $\Lambda$ + S picture are relaxed.  The resultant relatively short excited state lifetimes of the b0$^+$ and a2 states (15 $\mu$s and 2.4 ms, respectively, calculated using LEVEL 16~\cite{LeRoy2017}), are important for optical pumping schemes.  The ground $^3\Sigma ^-$ state is split into different $\Omega$ components separated by 1049 cm$^{-1}$, and we consider spectroscopy transitions within the $\Omega=0$ X$_10^+$ state.  We focus on the $^{130}$TeH$^+$ ion, whose lack of Te nuclear spin reduces the complexity of optical pumping. Optical pumping is further simplified because of the small reduced mass, leading to large rotational and vibrational constants predicted to be 6.2 and 2100 cm$^{-1}$, respectively. $^{130}$TeH$^+$ has $I=1/2$, and we use the Hund's case (c$_\beta$) basis such that $J_a=L+S$, $J=J_a+R$, and $F=J+I$.

An overtone transition from the $v=0$ to the $n$th vibrational state at frequency $\Omega_n$ has an absolute sensitivity to $\mu$ given by $\partial_\mu \Omega_n \equiv \frac{\partial \Omega_n}{\partial (\textrm{ln} \mu)}$.  For a molecule with fundamental vibrational frequency $\omega$, $\partial_\mu \Omega_n \approx n \omega/2$ in the harmonic region~\cite{Demille2008,Zelevinsky2008,Jansen2014,Hanneke2016}. Since $\partial_\mu \Omega$ is small for atomic transitions, $\partial_\mu \Omega_n$ with $n$ must eventually reverse toward dissociation; for a Morse potential the maximum occurs at $3/4$ the dissociation energy~\cite{Demille2008}.

What is the optimal overtone transition to use for a given molecule?  The statistical uncertainty in $\mu$ variation has been discussed for overtone measurements in homonuclear molecules~\cite{Hanneke2016}, where the natural linewidth $\Gamma$ of the transition is extremely narrow, and the maximum single-shot probe times are limited by other issues.  In contrast, vibrational state lifetimes of polar hydrides are sufficiently short (typically $< 1$ s) to limit probe times in realistic experiments.  Harmonic oscillator states provide an estimate for the natural lifetime $\tau_n$ of the $n$th state, valid for low $n$.  From $\bra{n-1} x \ket{n} \propto \sqrt{n}$ we obtain $\tau_n  \approx \tau_1/n$ in the harmonic region.

For a projection-noise limited Ramsey-style measurement of the RMS error for a single ion is given by:
\begin{equation}
\delta\Omega = \frac{1}{T_\textrm{R} C}\sqrt{\frac{T_\textrm{c}}{2 T}},
\label{precision}
\end{equation}
where $T_\textrm{R}$ is the Ramsey time, $T_\textrm{c}$ is the cycle time, $T$ is the total measurement time and $C$ is the fringe visibility (e.g. $C=0.6$ for $T_\textrm{R}=\tau_n$)~\cite{Riis2004, Hollberg2001}.
The experimentally obtainable statistical uncertainty for the overtone ($\delta\Omega_n$), which is generally worse than for the fundamental ($\delta\Omega_1$) because of the relative lifetimes, can be expressed as
\begin{equation}
\delta\Omega_n = \left(\frac{\tau_1}{\tau_n}\right)^k \delta\Omega_1,
\label{precision}
\end{equation}
with $0 \leq k \leq 1$.  The particular value of $k$ depends on the relationship between $T_\textrm{R}$, $T_\textrm{c}$, and $\tau$ in the measurement protocol. We consider three limiting cases: (1) $k=0$, for $T_\textrm{R}, T_\textrm{c} \ll \tau$, (2) $k=1$, for $T_\textrm{R}=\tau$ and $T_\textrm{R} \ll T_\textrm{c}$, and (3) $k=\frac{1}{2}$, for $T_\textrm{R}=\tau$ and $T_\textrm{C} = 2 T_\textrm{R}$.  The actual scenario for homonuclear vibrational spectroscopy corresponds to $k=0$, a dead-time limited experiment relevant to polar hydrides is described by $k=1$, and the best-case scenario for any given molecule is described by $k=\frac{1}{2}$.  Although it would be statistically preferable to have the longer state lifetimes of homonuclears, a sort of consolation for polar hydride spectroscopy is that their moderate lifetimes allow $k=\frac{1}{2}$ to be approached using optical pumping techniques~\cite{Stollenwerk2015}.

The statistical sensitivity $\zeta_n$ of the $\Omega_n$ transition to changing $\mu$ can then be defined and related to the sensitivity of the $\Omega_1$ transition as follows:
\begin{equation}
\zeta_n \equiv \frac{\partial_\mu \Omega_n}{\delta\Omega_n} =
\frac{\partial_\mu \Omega_n}{\partial_\mu \Omega_1} \left(\frac{\tau_n}{\tau_1}\right)^k \zeta_1.
\end{equation}
For a harmonic oscillator, the sensitivity and lifetime scalings discussed above yield (1) $\zeta_n / \zeta_1 = n$ for $k=0$, (2) $\zeta_n / \zeta_1 = 1$ for $k=1$, and (3) $\zeta_n / \zeta_1 = n^{1/2}$ for $k=1/2$.  The strongest benefit of increasing $n$ is seen for the $k=0$ case relevant to homonuclear spectroscopy.

\begin{figure}
\includegraphics[width=3.0in]{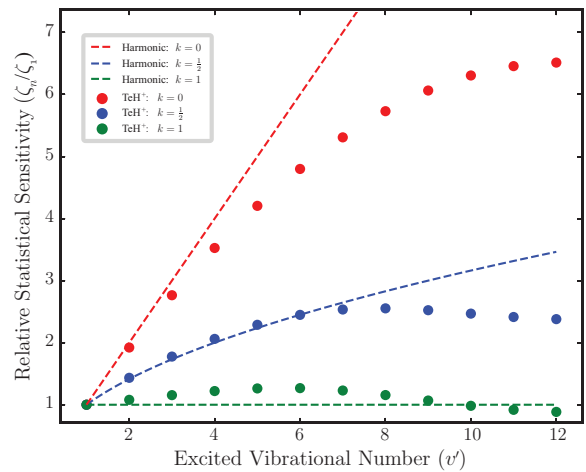}
\caption{\footnotesize Statistical sensitivity to $\mu$ of the $\Omega_n$ overtone transition, relative to that of the vibrational fundamental transition $\Omega_1$.  Values of $k$ correspond to different experimental timing cases, described in the text.}
\label{fig:sens}
\end{figure}

In TeH$^+$ the vibrational lifetimes in X$_10^+$ span from 200 ms down to 20 ms over the frequency range 60-600 THz. In Fig.~\ref{fig:sens} we plot the TeH$^+$ figure of merit for the three limiting cases of $k$, as a function of excited vibrational state.  Non-trivial dipole moment functions and reduced anharmonic level spacings contribute to the slight enhancement at low $n$ of the calculated FOM ratio, as compared with the harmonic oscillator values (also plotted).  The 15 $\mu$s lifetime of the diagonal X$10^+ - \textrm{b}0^+$ transition allows for rapid optical state preparation~\cite{Stollenwerk2015}, such that the linewidth-limited case $k=1/2$ can be approached.  In this case, the statistical uncertainty is minimized using the overtone $\Omega_8/(2\pi) = 430$ THz with $\tau_8= 40$ ms.  Averaging yields $\delta \Omega_8/\Omega_8 = 3.1 \times 10^{-15}$ $/\sqrt{T/\mathrm{sec}}$, or $1.0 \times 10^{-17}$ at one day, corresponding to a precision of 4.3 mHz. The absolute sensitivity of the $\Omega_8$ overtone to varying $\mu$ is $\partial_\mu \Omega_8 = 2 \pi \times 170$ THz.

Stark shifts, which arise from trap, blackbody, and laser electric fields, are important sources of systematic uncertainty.  For the normal case that the Zeeman interaction defines the quantization axis and for a linearly polarized field, the Stark shift can be expressed as
\begin{equation}
\Delta W = -\frac{1}{2}E_{\textrm{rms}}^2\left[\alpha_\textrm{S}(\omega)  + D\alpha_\textrm{T}(\omega)\frac{3M_J^2 - J(J + 1)}{(2J - 1)(2J + 3)} \right]
\label{eq:pol}
\end{equation}
where $E_\textrm{rms}$ is the rms value of the oscillating field polarized along $\hat{\textbf{u}}$, $D = (3|\hat{\textbf{u}}\cdot\hat{\textbf{z}}|^2-1)$, and $\alpha_\textrm{S}$ and $\alpha_\textrm{T}$ are the dynamic scalar and tensor polarizabilities~\cite{Brieger1984, itano_external-field_2000}.  (Note that we have defined $\alpha_\textrm{T}$ using the molecular convention, which causes the $M_J$-dependent factor to be $(2J)/(2J+3)$ times that of the atomic convention.) This expression has the correct form in the DC limit where $E_\textrm{DC}=E_\textrm{rms}$. If the rotational energy spacing is relatively small, then expressions of the same form as Eq.~\ref{eq:pol} can be written for polarizabilities arising from coupling to adjacent rotational levels ($\alpha^\textrm{r}$), adjacent vibrational manifolds ($\alpha^\textrm{v}$), the spin-orbit split $X_21$ manifold ($\alpha^\textrm{SO}$), and electronically excited manifolds ($\alpha^\textrm{e}$) (see Supplemental Material). Since the second term in Eq.~\ref{eq:pol} vanishes when summed over polarizations (as occurs naturally for an isotropic BBR field) or when measured spectra are averaged over Zeeman states, effects of the tensor polarizability can be strongly suppressed~\cite{dube_electric_2005, dube_evaluation_2013}.  The scalar polarizabilities are of greater concern.

To anticipate the magnitude of electronic Stark shifts from Fig.~\ref{fig:pec}, it is important to recognize that the vibrational wavefunctions generally cause the upper spectroscopy state to couple to other electronic manifolds well above their minimum energies.  We estimate the polarizability arising from  the A$_2$1 state using a `classical' approximation~\cite{lawley_photodissociation_2009,tellinghuisen_resolution_1983, tellinghuisen_efficiency_1984} described in the Supplemental Material, with results in Table~\ref{tab:pol}.  Transition moments to other nearby electronic states are small, and they are not expected to contribute significantly to the polarizability.  Polarizabilities arising from coupling to different spin-orbit states are also small, owing to small transition moments.

Because the rotational and vibrational spacing is much smaller than the electronic spacing,  $\alpha^\textrm{v}(\omega)$ or $\alpha^\textrm{r}(\omega)$ for polar molecules might be expected to dominate the differential Stark shift at low frequencies and also to play a significant role at high frequencies.  For the case of $\alpha^\textrm{v}(\omega)$, this turns out not to be the typical case for straightforward reasons.  Vibrational transition moments for polar hydrides are typically $\le 10\%$ of electronic transition moments, so after squaring to obtain the Stark shift, vibrational contributions to Stark shifts are still smaller than electronic contributions despite the closer level proximity.  The case of $\alpha^\textrm{r}(\omega)$ is much more interesting and is discussed below.

We find that the relationship between the rigid rotor level spacings and corresponding dipole matrix elements essentially eliminates effects of $\alpha_\textrm{S}^\textrm{r}$. In the low frequency limit, the shifts from the next-lower and next-higher rotational states balance, and $\alpha_\textrm{S}^\textrm{r}(0) = 0$ for $J \ge 1$.  This result is consistent with well-known expressions for DC Stark shifts of a rigid rotor~\cite{kronig_dielectric_1926,Brown2003}, with the observation that the $M_J$-averaged DC polarizability vanishes for $J>0$~\cite{Cheng2007}, and with calculations showing that $\alpha_\textrm{S}(0)$ for HD$^+$ is dominated by $\alpha_\textrm{S}^\textrm{e}(0)$~\cite{schiller_static_2014}. In the Supplemental Material, we prove these results for Hund's case (c$_\beta$) which applies to TeH$^+$; we have also found that they apply to Hund's case (a) and case (b).  Centrifugal distortion has a small affect on the rotational spacing ($<10^{-4}$ per level for \TeH) and slightly spoils this cancellation, as can be seen in Table~\ref{tab:pol}.   We also find in the high-frequency limit that all levels obtain a common $\alpha_\textrm{S}^\textrm{r}$, which can be seen in the rotational contributions to BBR shifts in Table~\ref{tab:pol}.  However, this is a less significant result for \TeH\ since the differential electronic BBR shift is relatively large.  BBR shifts were calculated by numerically integrating the dynamic Stark shifts over the BBR spectrum~\cite{Farley1981}.

\begin{table*}
\begin{ruledtabular}
\begin{tabular}{l | l l l l | l l l l | l l l l }
 &\multicolumn{4}{c}{$\alpha_\textrm{S}(\omega=0)$ (a.u.)}    &\multicolumn{4}{c}{$\alpha_\textrm{T}(\omega=0)$ (a.u.)} &\multicolumn{4}{c}{$\Delta_{300}$ (mHz)}\\
\hline
 & $\alpha^\textrm{r}_\textrm{S}$ & $\alpha^\textrm{v}_\textrm{S}$ & $\alpha^\textrm{so}_\textrm{S}$ & $\alpha^\textrm{e}_\textrm{S}$ &  $\alpha^\mathrm{r}_\mathrm{T}$ & $\alpha^\textrm{v}_\textrm{T}$ & $\alpha^\textrm{so}_\textrm{T}$ & $\alpha^\textrm{e}_\textrm{T}$ &$\Delta^\textrm{r}_{300}$
 &$\Delta^\textrm{v}_{300}$ &$\Delta^\textrm{SO}_{300}$ &$\Delta^\textrm{e}_{300}$\\
\hline
\ket{0,0} &1500 &0.02 &0.04 &1 &0 &0 &0 &0 &12 &-0.2 &-0.4 &-10   \\
\ket{0,1}$^*$ &0.08 &0.02 &0.04 &1 &1100 &-0.04 &0.02 &0.6 &11 &-0.2 &-0.2 &-10  \\
\ket{0,2} &0.2 &0.02 &0.04 &1 &400 &-0.04 &0.02 &0.6 &11 &-0.2 &-0.2 &-10  \\
\hline
\ket{8,2}$^*$ &0.3 &-0.03 &0.03 &0.6 &600 &-0.09 &0.01 &0.3 &11 &0.4 &-0.1 &-6 \\
\end{tabular}
\end{ruledtabular}
\caption{\footnotesize Contributions to scalar and tensor DC polarizabilities and 300 K BBR shifts for selected $X_10^+$ $J$-states $\ket{v, J}$.  Tensor polarizabilities use the molecular convention (see Supplemental Material).  The proposed spectroscopy transition is marked$^*$.}
\label{tab:pol}
\end{table*}

In Table~\ref{tab:comparison} it is seen that the differential scalar polarizabilities and BBR shifts of TeH$^+$ compare favorably with those of atoms. The dominant electronic dipole transition moments in molecules are typically a few times smaller than those in atoms, so it is not surprising that molecular electronic polarizabilities compare favorably. Apart from vanishing for $J<1$ (or $F<1$) states, the molecular $\alpha_\textrm{T}^\textrm{r}(0)$ is generally large but can be dealt with by averaging over Zeeman levels.

When driving a relatively weak overtone transition to an upper state where stronger decay channels are open, the light shift from the spectroscopy laser must be considered.  The saturation intensity $I_\textrm{sat} \propto \Gamma^2/\mu_\textrm{eg}^2$, where $\Gamma$ is the total relaxation rate and $\mu_\textrm{eg}$ is the spectroscopy interval transition moment.   Contrary to the two-level case, saturating a weaker (higher) overtone transition $v=0 \rightarrow n$ requires increased intensity, since $\Gamma$ increases with $n$ but $\mu_\textrm{eg}$ decreases. For the \TeH\ $v=0 \rightarrow 8$ transition, the upper state has $\Gamma =  25\ \textrm{s}^{-1}$, and the spectroscopy channel has $\Gamma_{80} = 2.4 \times 10^{-4}$ s$^{-1}$, yielding $I_\textrm{sat}= 1.5\ \mu\textrm{W/mm}^2$.  At this drive intensity, the estimated differential light shift is 0.5 mHz (a fractional shift of $1 \times 10^{-18}$), dominated by coupling of $v=8$ to the A$_2$1 state.  Spectroscopy laser intensity and pointing control can stabilize the shift to below this level.

When the hyperfine structure is considered, it at first glance appears attractive to perform spectroscopy on $F=1/2$ components of a $J=0 \rightarrow J=1$ transition, since for these states there is no quadrupole or tensorial Stark shifts.  However, the large scalar polarizability of $J=0$  makes this transition problematic, and unlike quadrupole and tensorial Stark shifts, it cannot be reduced by simple averaging over Zeeman levels.  Additionally, quadratic Zeeman shifts cannot be mitigated by similar averaging.  To choose promising spectroscopy states, we diagonalize the effective Hamiltonian described in the Supplemental Material.

\begin{table*}
\begin{ruledtabular}
\begin{tabular}{l | l l l l l l l l}
 &$\Delta \alpha_\textrm{S}(0)$ (a.u.) &$\Delta \alpha^\textrm{(a)}_\textrm{T}(0)$ (a.u.) &$\Delta f_{300}$ (Hz) &$g_\textrm{g}$ &$g_\textrm{e}$ &$\Delta_\textrm{M2}$ (Hz/mT$^2$) &$\Delta \Theta$ (a.u.) &$\delta f/f \times 10^{18}$     \\
 \hline
TeH$^+$ (430 THz) &-0.4 &250 &0.005 &0.07 &0.05 &40,000 &0.3 &10 \\
Al$^+$ (1100 THz) &0.5 &0 &-0.004 &-0.0008  &-0.002 &-70 &0 &0.3$^*$ \\
Sr$^+$ (445 THz) &-30 &-50 &0.2  &2 &1 &3 &3 &3 \\
Yb$^+$ E2 (688 THz) &50 &-70 &-0.4  &- &0.8 &50,000 &2 &6 \\
Yb$^+$ E3 (642 THz) &5 &-1 &-0.07  &- &1 &-2000 &-0.04 &0.6$^*$ \\
\end{tabular}
\end{ruledtabular}
\caption{\footnotesize Comparison of \TeH\ $\ket{v=0,J=1,F=1/2} \rightarrow \ket{v=8,J=2,F=3/2}$ and atomic ion clock transition parameters~\cite{Ludlow2015, schneider_sub-hertz_2005,Huntemann2016,baynham_measurement_2018}.  Differential shift coefficients are given for DC polarizabilities $\Delta \alpha(0)$, 300 K BBR Stark shift $\Delta f_{300}$, quadratic Zeeman shift $\Delta_\textrm{M2}$, and quadrupole shift $\Delta \Theta$. For comparison tensor polarizabilities here use the atomic convention, denoted $\alpha^\textrm{(a)}_\textrm{T}$, so \TeH\ values are smaller than those of Table~\ref{tab:pol} by a factor of $2J/(2J+3)$.  $\Delta \alpha$ and $\Delta \Theta$ are the differences between the upper and lower state values. Lower and upper state linear Zeeman shifts are given by the g-factors $g_g$ and $g_e$, where $E = g m_F \mu_B$.  The quadratic Zeeman coefficient $\Delta_\textrm{M2}$ is either for $m_F=0 \rightarrow m_F'=0$ or for an average over Zeeman components effectively creating that transition.
The statistical uncertainty $\delta f/f$ is for 1 day of averaging, with $T_\textrm{R}$ set to the upper state lifetime for \TeH\ and Sr$^+$ and $T_\textrm{R}$ $^*$set to a laser coherence time of 6~s~\cite{bishof_optical_2013} for Al$^+$, and Yb$^+$.}
\label{tab:comparison}
\end{table*}

Spectroscopy states within the X$_10^+$ manifold have intrinsically small linear Zeeman shifts, due to a lack of electronic angular momentum. The remaining moments are of order a nuclear magneton. However, X$_10^+$ acquires some electronic spin via its rotational-electronic coupling with X$_21$. This type of mixing, also sometimes called Coriolis coupling, can sometimes significantly affect the spectrum~\cite{carrington_microwave_1995, McGuyer2013, Borkowski2014}.
Since the $\Omega$-doubling in X$_21$ is primarily caused by rotational-electronic coupling to nearby electronic states of $\Omega = \pm 1$, it can be used to estimate the degree of mixing.  The resulting $X_10^+$ magnetic moments are of order 0.1 $\mu_B$, shown in Table~\ref{tab:comparison}. First order shifts can be mitigated by averaging over pairs of transitions with opposite $M_F$ yielding an average frequency with no linear shift~\cite{Nicholson2015}.  This technique has achieved suppression of Bohr-magneton sized linear Zeeman shifts at the $<10^{-17}$ level~\cite{Dube2015}, leading us to project an uncertainty at $<10^{-18}$. Alternatively , the first order Zeeman shifts could be reduced by probing $M_F = 0 \rightarrow M_F' =0$ transitions in $^{125}$TeH$^+$, which will also have relatively small quadratic Zeeman shifts due to larger hyperfine splitting~\cite{Fujiwara1997}.

Quadratic Zeeman shifts arise from $M_F$-preserving mixing between hyperfine states $F=J \pm I$. Although the $J=1$ and $J=2$ manifolds each have a pair of stretched states with $|M_F|=F=J+1/2$ possessing small quadratic Zeeman shifts, an alternative would allow nulling of quadrupole shift and tensorial Stark shifts.  We propose averaging over four spectroscopy transitions: $\ket{v=0,J=1,F=1/2,M_F=1/2} \rightarrow \ket{v=8,J=2,F=3/2, M_F=1/2 (3/2)}$ and their negative $M_F$ partners. These transitions have smaller differential quadratic Zeeman shifts than would transitions involving a stretched state. This state averaging protocol is similar to schemes effectively implemented in atoms to null these shifts~\cite{dube_electric_2005, dube_evaluation_2013}.

The calculated quadrupole moment function~\cite{OrnellasPersonal2018} yields $\Theta = 0.54\ e a_0^2$ for \ket{X_10^+, v=8, J=1, F=3/2, M_F'}, similar to typical values for atoms.  The simple averaging protocol discussed above can be used to effectively eliminate this shift.


\begin{table}
\begin{ruledtabular}
\begin{tabular}{l c}
Effect &$\sigma/f \times 10^{18}$ \\
\hline
BBR Stark & 0.9 \\
DC Stark, Scalar & 0.09 \\
DC Stark, Tensor & $\ll 1$ \\
Light shift & $<1$ \\
Quadrupole & $<1$ \\
Quadratic Zeeman & 0.6 \\
Statistics (at 1 day) & 10 \\
\end{tabular}
\end{ruledtabular}
\caption{\footnotesize Projected uncertainty for spectroscopy on \TeH\ $\ket{v=0, J=1, F=1/2} \rightarrow \ket{v=8, J=2, F=3/2}$.}
\label{tab:projections}
\end{table}

Projected limits to experimental precision are given in Table~\ref{tab:projections}.  Accuracy in some cases could be worse than precision, e.g. if an absolute shift is not well known, but holding shifts stable is sufficient for measuring varying $\mu$.  Values are obtained for a bias field of 300 nT, more than sufficient to resolve the Zeeman components, and a field uncertainty of 10 nT, which is a few times worse than achieved in~\cite{Madej2012}.  We use an electric field uncertainty of 100 V/m which is not the best achieved in single-ion experiments~\cite{keller_precise_2015} but is similar to the level arising in a 2-ion experiment where an uncompensated 10 V/m DC field~\cite{rosenband_frequency_2008} pushes the ions off-axis into a finite rf field.  We assume a suppression of the tensorial Stark shift by the $M_F$ averaging techniques discussed above; a suppression by 1000 would make these shifts similar in magnitude to the scalar Stark shifts.  The BBR uncertainty is from a 5 K temperature stability at 300 K.  For the tensorial Stark, quadrupole, and linear Zeeman shifts, we have assumed averaging over Zeeman sublevels as discussed above, which has been demonstrated in atomic clocks to strongly suppress uncertainties.

In conclusion, we have demonstrated the potential for single-photon vibrational overtone spectroscopy on a single polar molecular ion to reach systematic uncertainties at the $10^{-18}$ level and statistical uncertainties at the $10^{-17}$ level for one day of averaging.  Using the absolute sensitivity $\partial_\mu \Omega_8 = 2 \pi \times 170$ THz of the \TeH\ $\Omega_8$ overtone transition at 430 THz, we conclude that taking measurements over the course of a year could probe for varying $\mu$ with a sensitivity approaching the $1 \times 10^{-18}/\textrm{yr}$ level.

This small systematic uncertainty comes as somewhat of a surprise, since polar molecules have closely spaced rotational levels which can mix and cause large Stark shifts for low frequency fields.  In this work we point out that the associated polarizability is scalar in character for $J=0$ and indeed a significant issue, but that it is tensorial in character for $J>0$ and can thus be mitigated by simple averaging protocols regularly used in atomic clocks.  The vanishing of this $J>0$ DC scalar polarizability arises from a fortuitous relationship between rigid rotor oscillator strengths and level spacings.

Our results suggest that atoms, polar, and nonpolar molecules can reach similar levels of systematic uncertainty, e.g. they all have electronic polarizabilities which ultimately determine Stark shift uncertainties.  However, statistical uncertainties are expected to be quite different.  Homonuclear vibrational state lifetimes are much longer than polar lifetimes, but the actual statistical uncertainty will depend heavily on details of the experimental cycle, such as state preparation time.

Statistical uncertainty will ultimately limit the reach of single-ion spectroscopy on \TeH.  To improve the statistical reach of this proposal, the isotopologue TeD$^+$ is of interest because it is predicted to have overtone linewidths twice as narrow.  The relatively short 15 $\mu$s lifetime of the \TeH\ diagonal b0$^+$-X$0^+$ transition might allow fluorescence state readout of multiple ions~\cite{Arnold2015, arnold_suppression_2016, Keller2015}. Performing spectroscopy on a $\ket{J=0,F=1/2} \rightarrow \ket{J=1,F=1/2}$ transition would give the ions a negative, albeit large, $\alpha_\textrm{S}$ which might allow precision spectroscopy on a 2D or 3D crystal with the rf frequency properly tuned such that the Stark and second-order Doppler shifts cancel~\cite{Arnold2015}.  This transition would also be free of tensorial Stark and quadrupole shifts. Finally, we note that the vibrational state lifetimes of \TeH\ are not particularly long compared with other polar species (e.g. a $v=1$ lifetime of 4.0 s in CD$^+$~\cite{ornellas_theoretical_1986} as compared with 0.2 s in \TeH), so searching for other coolable candidates with favorable properties is well motivated.

\begin{acknowledgements}
This work was supported by AFOSR Grant No. FA9550-13-1-0116, NSF Grant No. PHY-1404455, and NSF GRFP DGE-1324585. We appreciate computational data on \TeH\ theory provided by Antonio Gustavo S. de Oliveira-Filho and  Fernando R. Ornellas.  We thank James Chou and Stephan Schiller for stimulating conversations.
\end{acknowledgements}

\widetext
\clearpage
\begin{center}
\textbf{\large Supplemental Material: Prospects for Polar Molecular Ion Optical Probe of Varying Proton-Electron Mass Ratio}
\end{center}
\setcounter{equation}{0}
\setcounter{table}{0}
\setcounter{page}{1}
\makeatletter
\renewcommand{\theequation}{S\arabic{equation}}

\section{Polarizabilities}
Our main results do not actually require a discussion of polarizabilities.  We are interested in the Stark shifts and the $M_F$-averaged Stark shifts, all of which we actually compute directly from the Hamiltonian rather than using polarizabilities.  However, since we find that some shifts vanish when averaged over $M_F$, a description in terms of scalar and tensor polarizabilities is suggested, and this description is also helpful for comparing behavior of different species.

\subsection{Polarizabilities for $J$-States}
Eq. (3) from the main text can be broken down into contributions to Stark couplings to the various manifolds, \begin{equation}
\Delta W = \Delta W^\textrm{r} + \Delta W^\textrm{SO} + \Delta W^\textrm{v} + \Delta W^\textrm{e}.
\end{equation}
In the approximation that (1) the rotational spacing is small compared with other intervals and that (2) electronic and vibrational transition dipole moments do not change significantly when the rovibrational wavefunction $v(J)$ is replaced with $v(J')$ for $J'=J\pm1$~\cite{Brieger1984}, then the (orientation-dependent) individual terms for $\Omega=0$ states can be written as
\begin{equation}
\Delta W^{x}(\gamma, J, M_J) = -\frac{1}{2}E_{\textrm{rms}}^2\left[\alpha_\textrm{S}^x(\gamma, J; \omega)  + D\alpha_\textrm{T}^x(\gamma, J; \omega)\frac{3M_J^2 - J(J + 1)}{(2J - 1)(2J + 3)} \right],
\label{eq:pol}
\end{equation}
with $D = (3|\textbf{u}\cdot\textbf{z}|^2-1)$ and $x \in \{$r, SO, v, e$\}$.  We have used the standard convention for defining the molecular $\alpha_\textrm{T}$, in which the $M_J$-dependent multiplier in Eq.~\ref{eq:pol} is $(2J)/(2J+3)$ times that of the atomic convention.  With these definitions, Ref.~\cite{Brieger1984} shows that for $I=0$ the electronic polarizabilities have the following relations:
\begin{equation}
\begin{split}
& \alpha_\textrm{S}^\textrm{e} = \frac{1}{3}\left[\alpha_\parallel + 2\alpha_\perp \right], \\
& \alpha_\textrm{T}^\textrm{e} = \frac{2}{3}\left[\alpha_\parallel - \alpha_\perp \right].
\end{split}
\label{parperp}
\end{equation}
Here, $\alpha_\parallel$ and $\alpha_\perp$ are the electronic polarizabilities associated with $\Delta\Omega = 0$ and $\Delta\Omega \pm 1$ transitions, and ($\alpha_\parallel - \alpha_\perp$) is known as the polarizability anisotropy.  Since understanding the general Stark shift properties of the molecule does not require introducing nuclear spin, and since its introduction complicates the connection with the polarizability anisotropy, we report polarizabilities for $J$ states in Table I of the main text.

\subsection{Determining $J$-State Scalar and Tensor Polarizabilities}
The second-order perturbation expression for the Stark shifts of a state \ket{\gamma, J, M_J} coupled to a manifold $\gamma'$ by an oscillating electric field $\textbf{E}(t)=\mathcal{E} \cos{\omega t}\ \hat{\textbf{z}}$ is given by
\begin{equation}
\Delta W^{x}(\gamma, J, M_J) = \sum_{J'}\frac{E_\textrm{rms}^2 |\bra{\gamma, J, M_J}d_z\ket{\gamma',J',M_J}|^2}{\hbar} \frac{-\Delta}{\Delta^2 - \omega^2},
\label{eq:shift}
\end{equation}
where $d_z$ is the lab-frame z-component of the dipole operator, $\hbar \Delta$ is the signed energy splitting of the states, and $E_\textrm{rms}$ is the rms field magnitude.  In this work, we find the polarizabilities by diagonalizing the Hamiltonian. Combining Eqs.~\ref{eq:pol} and~\ref{eq:shift}, and recognizing that the tensorial term vanishes when summed over all $M_J$, we obtain
\begin{equation}
\alpha^x_S(\gamma, J; \omega) = \sum_{M_J} \sum_{J'} \frac{2 |\bra{\gamma, J, M_J}d_z\ket{\gamma',J',M_J}|^2}{\hbar (2J+1)}\frac{\Delta}{\Delta^2 - \omega^2}.
\label{eq:scalar}
\end{equation}
The tensor polarizability $\alpha_\textrm{T}^x$ for the manifold can then be found from Eq.~\ref{eq:pol}.  For instance, choosing $M_J=0$ and $\textbf{u}=\textbf{z}$ such that $D=2$ we obtain
\begin{equation}
\alpha^x_T(\gamma, J; \omega) = \frac{(2J-1)(2J+3)}{2J(J+1)} \left[-\alpha^x_S(\gamma, J; \omega)+
\sum_{J'} \frac{2 |\bra{\gamma, J, 0} d_z \ket{\gamma',J',0}|^2}{\hbar}\frac{\Delta}{\Delta^2 - \omega^2}\right].
\label{eq:tensor}
\end{equation}

\subsection{Rotational Polarizability}
The level spacing for a rigid rotor is $E_J = B_v J (J+1)$, yielding an upper energy interval
\begin{equation}
\Delta_{J \rightarrow J+1} = 2 (J+1) B_v / \hbar,
\end{equation}
and a signed downward interval
\begin{equation}
\Delta_{J \rightarrow J-1} = -2 J B_v / \hbar.
\end{equation}

For a Hund's case (c) molecule with body-frame dipole moment $\mu_0$ in a z-polarized field, the polarizability due to coupling to adjacent rotational levels, from Eq.~\ref{eq:pol} and e.g.~\cite{Brown2003}, becomes
\begin{align}
\alpha^\textrm{r}_S(\omega)  &= \sum_{M_J} \sum_{J'} \frac{2 |\bra{\gamma,J, M_J} - \mu_z \ket{\gamma, J', M_J}|^2}{\hbar(2J+1)} \frac{\Delta_{JJ'}}{\Delta_{JJ'}^2 - \omega^2} \\
&=2 \mu_0^2 \sum_{M_J} \sum_{J'}(2J'+1)  \left|\tj{J}{1}{J'}{-M_J}{0}{M_J} \tj{J}{1}{J'}{-\Omega}{0}{\Omega}\right|^2 \frac{\Delta_{JJ'}}{\Delta_{JJ'}^2 - \omega^2}
\end{align}
We use
\begin{equation}
\tj{J+1}{1}{J}{-M_J}{0}{M_J} = (-1)^{J-M_J+1}\left[\frac{2(J+M_J+1)(J-M_J+1)}{(2J+1)(2J+2)(2J+3)}\right]^{1/2}
\end{equation}
and for $\omega \ll \Delta$
\begin{equation}
\frac{\Delta_{JJ'}}{\Delta_{JJ'}^2 - \omega^2} \approx \frac{1}{\Delta} \left(1+\frac{\omega^2}{\Delta^2}\right).
\end{equation}
Taking $\Omega=0$, $J'=1$ for $J=0$ or $J'=J \pm 1$ for $J \ge 1$, for the low frequency limit $\omega \ll \Delta$ we obtain
\begin{equation}
\alpha^\textrm{r}_S(\omega) \approx
\left\{
\begin{aligned}
&\left(\frac{\mu_0^2}{3 B_v}\right) \left[1+\left(\frac{\hbar \omega}{2 B_v}\right)^2\right] &\xrightarrow[\omega \rightarrow 0]{}& \quad  \frac{\mu_0^2}{3 B_v}, &J=0 \\
&\left(\frac{\mu_0^2}{3 B_v}\right) \left(\frac{1}{J(J+1)}\right)^2 \left(\frac{\hbar \omega}{2 B_v}\right)^2 &\xrightarrow[\omega \rightarrow 0]{}& \quad  0, &J \ge 1
\end{aligned}
\right.
\end{equation}
For $J \ge 1$ states, the cancellation to zeroth order occurs because the interactions with the next-lower and next-higher states cancel each other.  This is a non-trivial result, since the level spacing and coupling strengths are different for each interval. This result is consistent with the well known DC Stark shifts~\cite{kronig_dielectric_1926,Brown2003} which were also reported as DC polarizabilities in~\cite{Cheng2007}.

One can also show that in the limit $\omega \gg B_v/\hbar$ that rotational coupling causes all levels except for $J=1/2$ to obtain a common scalar polarizability $\alpha_\textrm{S} = -\frac{4 \mu_0^2 B_v}{3 \hbar^2 \omega^2}$.  This result is relevant, for instance for Stark shifts from BBR coupling of rotational levels.  However, in practice since these shifts are small, this cancellation is less important than the low-frequency case.

\subsection{Polarizability From an Unbound Electronic State}
For computing the polarizability from coupling to the unbound A$_2$1 level, we use a `classical' approximation which takes the classical position and momentum of the spectroscopy state as a function of internuclear distance $R$ and uses the requirement of conservation of nuclear position and momentum to define a single energy in the excited potential which is coupled.  It is easily shown that the coupling interval is given by the so-called difference potential $\Delta V(R)$, the interval between the two potential energy curves~\cite{tellinghuisen_joel_franckcondon_2007}. Then the frequency shift in response to an off-resonant field is given by
\begin{equation}
\Delta f = \int dR \Delta f(\mu(R), \Delta V(R), E(t)) |\Psi(R)|^2,
\end{equation}
where $\Psi(R)$ is the nuclear wavefunction, $\mu(R)$ is the lab-frame transition moment, and $\Delta f(\mu, \Delta E_, E(t))$ describes the light shift for driving field $E(t)$ of a level coupled by transition moment $\mu$ to another level separated in energy $\Delta E$.  The results we obtain by integrating over $ |\Psi(R)|^2$ are similar to what we obtain by a simple turning point approximation. Lab frame transition moments are obtained from rotationless transition moments in the usual way, the shifts are summed over coupled $F', M_F'$ levels, and scalar and tensorial polarizabilities are extracted as described previously.

\subsection{Polarizabilities for $F$-States}
When comparing spectroscopy transition differential polarizabilities with those of atoms in Table II, it is important to consider the $F$ states.  It can be shown for atoms that the polarizability for $F$ states can be written in the same form as Eq. 3 of the main text, but with $J \rightarrow F$ and $M_J \rightarrow M_F$~\cite{itano_external-field_2000}.  Since we are considering Hund's case (c$_\beta$) where $F=J + I$, the same arguments apply to TeH$^+$.  In order to report values in Table II, we calculate the numerical Stark shifts and solve the equation for $\alpha_\textrm{S}(\gamma,F;\omega)$ and $\alpha_\textrm{T}(\gamma,F;\omega)$.  For the sake of comparison, the prefactor for $\alpha_\textrm{T}$ given in this table follows the atomic convention.

\section{Zeeman Shifts}
The matrix elements of the Hamiltonian are adapted from~\cite{Brown2003} and described below.

The effective Hamiltonian for the X$_10^+$ and X$_21$ states are
\begin{equation} \label{Hamiltonian}
\mathcal{H} =  \mathcal{H}_{\mathrm{rot}} + \mathcal{H}_{\mathrm{nsr}} + \mathcal{H}_{\mathrm{HFS}} + \mathcal{H}_{Z_I} + \mathcal{H}_{Z_\mathrm{rot}} + \mathcal{H}_{Z_\mathrm{S}} + \mathcal{H}_{E} + \mathcal{H}_{Q}
\end{equation}

where $\mathcal{H}_{\mathrm{rot}}$ is the rotational kinetic energy and

\begin{equation} \label{Nuclear_Spin_Rotation}
\mathcal{H}_{\mathrm{nsr}} =  -c_I T^1(\bm{J}) \cdot T^1(\bm{I}),
\end{equation}

\begin{equation} \label{Hyperfine}
\mathcal{H}_{\mathrm{HFS}} =  b T^1(\bm{S}) \cdot T^1(\bm{I}) + c T^1_{q=0}(\bm{S}) \cdot T_{q=0}^1(\bm{I}),
\end{equation}

\begin{equation} \label{Zeeman_I}
\mathcal{H}_{Z_I} =  -g_I \mu_N T^1_0(\bm{B}) \cdot T^1_0(\bm{I}),
\end{equation}

\begin{equation} \label{Zeeman_Rotation}
\mathcal{H}_{Z_\mathrm{rot}} =  -g_J \mu_B T^1_0(\bm{B}) \cdot T^1_0(\bm{J}),
\end{equation}

\begin{equation} \label{Zeeman_S}
\mathcal{H}_{Z_\mathrm{S}} =  g_s \mu_B T^1_0(\bm{B}) \cdot T^1_0(\bm{S}),
\end{equation}

\begin{equation} \label{Stark}
\mathcal{H}_E =  -T^1_0 ( \bm{\mu_e} ) \cdot T^1_0(\bm{E}),
\end{equation}

\begin{equation} \label{Quadrupole}
\mathcal{H}_Q =  -T^2_0 ( \bm{\nabla E} ) \cdot T_0^2(\bm{Q}).
\end{equation}

The constants $c_I, g_I, g_J, g_s$ and $\mu_e$ are the nuclear spin-rotation coupling constant, proton g factor, rotational g factor, electron spin g factor and ground state body-frame electric dipole moment, respectively. The values are presented in Table~\ref{table:constants}.  The $I \cdot L$ and $B \cdot L$ terms are omitted because $L = 0$ for the X states, and $L$ can be considered a good quantum number because the pure-precession hypothesis is well justified for hydrides~\cite{Brown2003}. For convenience, the diagonal and off-diagonal components of the rotational Hamiltonian, $\mathcal{H}_{\mathrm{rot}}^D$ and $\mathcal{H}_{\mathrm{rot}}^{OD}$, are separated, where $\mathcal{H}_{\mathrm{rot}} = \mathcal{H}_{\mathrm{rot}}^D + \mathcal{H}_{\mathrm{rot}}^{OD}$.  The magnetic field along the laboratory $\hat{z}$ axis defines the quantization direction.

\begin{equation}
\begin{aligned}
&\langle v, J_a; \Omega, J|\mathcal{H}_{\mathrm{rot}}^D|v, J_a; \Omega, J\rangle = B_v\left[J(J + 1) + J_a(J_a + 1) - 2\Omega^2 \right] \\[10 pt]
&\langle v, J_a; \Omega, J|\mathcal{H}_{\mathrm{rot}}^{OD}|v, J_a; \Omega', J\rangle = -2B_v \sum_{q = \pm 1} (-1)^{J_a - \Omega}
\begin{pmatrix}
    J_a & 1 & J_a \\
    -\Omega & q & \Omega'
\end{pmatrix}
 (-1)^{J - \Omega}
\begin{pmatrix}
    J & 1 & J \\
    -\Omega & q & \Omega'
\end{pmatrix} \\
&\qquad \qquad \times \left[J_a(J_a + 1)(2J_a + 1)J(J + 1)(2J + 1)\right]^{1/2}\\
&\qquad \qquad \textrm{where }q=\Omega-\Omega'\\[10 pt]
&\langle v,~ \Omega, J, I, F|\mathcal{H}_{\mathrm{nsr}}|v, \Omega, J, I, F\rangle
= c_I  (-1)^{J + F + I}
\begin{Bmatrix}
    I & J & F \\
    J & I & 1
\end{Bmatrix} \\
&\qquad \qquad \times [J(J + 1)(2J + 1)I(I + 1)(2I + 1)]^{1/2} \\[10 pt]
&\langle v, J_a; \Omega, J, I, F|\mathcal{H}_{\mathrm{HFS}}|v, J_a; \Omega', J', I, F'\rangle \\
&\qquad \qquad = ~x(-1)^{J' + F + I}
\begin{Bmatrix}
    I & J' & F \\
    J & I & 1
\end{Bmatrix}
[I(I + 1)(2I + 1)]^{1/2}
  (-1)^{J - \Omega}
\begin{pmatrix}
    J & 1 & J' \\
    -\Omega & q & \Omega'
\end{pmatrix} \\
& \qquad \qquad \times (-1)^{J_a - \Omega}
\begin{pmatrix}
    J_a & 1 & J_a \\
    -\Omega & q & \Omega'
\end{pmatrix}
(-1)^{J_a + L + S + 1}
\begin{Bmatrix}
    J_a & S & L \\
    S & J_a & 1
\end{Bmatrix} \\
& \qquad \qquad \times[(2J_a' + 1)(2J_a + 1)S(S + 1)(2S + 1)]^{1/2} \\
& \qquad \qquad \textrm{where }x = b,c. \\[10 pt]
&\langle v,~ \Omega, J, I, F, M_F|\mathcal{H}_{Z_\mathrm{rot}}|v, \Omega, J, I, F', M_F\rangle \\
& \qquad \qquad = -g_J \mu_B B_z (-1)^{F - M_F}
\begin{pmatrix}
    F & 1 & F' \\
    -M_F & 0 & M_F
\end{pmatrix}
(-1)^{F' + J + 1 + I}[(2F' + 1)(2F + 1)]^{1/2} \\
& \qquad \qquad \times
\begin{Bmatrix}
    J & F' & I \\
    F & J & 1
\end{Bmatrix}
[J(J + 1)(2J + 1)]^{1/2} \\[20pt]
&\langle v,~  \Omega, J, I, F, M_F|\mathcal{H}_{Z_I}|v, \Omega, J, I, F', M_F\rangle \\
&\qquad \qquad = -g_I \mu_N B_z (-1)^{F - M_F}
\begin{pmatrix}
    F & 1 & F' \\
    -M_F & 0 & M_F
\end{pmatrix}
(-1)^{F + J + 1 + I}[(2F' + 1)(2F + 1)]^{1/2} \\
& \qquad \qquad \times
\begin{Bmatrix}
    F & I & J \\
    I & F' & 1
\end{Bmatrix}
[I(I + 1)(2I + 1)]^{1/2} \\[10 pt]
&\langle v, J_a; \Omega, J, I, F, M_F|\mathcal{H}_{Z_S}|v, J_a; \Omega', J', I, F', M_F\rangle \\
& \qquad \qquad = ~g_s \mu_B B_z (-1)^{F - M_F + F' + 2J + I + 1 - \Omega}
\begin{pmatrix}
    F & 1 & F' \\
    -M_F & 0 & M_F
\end{pmatrix}
\begin{pmatrix}
    J & 1 & J' \\
    -\Omega & q & \Omega'
\end{pmatrix}
[(2F' + 1)(2F + 1)]^{1/2} \\
  &\qquad \qquad \times [(2J' + 1)(2J + 1)]^{1/2}
\begin{Bmatrix}
    J & F & I \\
    F' & J' & 1
\end{Bmatrix}
(-1)^{J_a - \Omega}
\begin{pmatrix}
    J_a & 1 & J_a \\
    -\Omega & q & \Omega'
\end{pmatrix}
(-1)^{J_a + L + S + 1} \\
&\qquad \qquad \times
\begin{Bmatrix}
    J_a & S & L \\
    S & J_a & 1
\end{Bmatrix}
[(2J_a' + 1)(2J_a + 1)S(S + 1)(2S + 1)]^{1/2}\\[10 pt]
&\langle v,~ \Omega, J, I, F, M_F|\mathcal{H}_{E}|v, \Omega, J', I, F', M_F\rangle \\
&\qquad \qquad = -\mu_e E_0 (-1)^p (-1)^{F - M_F}
\begin{pmatrix}
    F & 1 & F' \\
    -M_F & p & M_F'
\end{pmatrix}
(-1)^{F' + J + 1 + I}[(2F' + 1)(2F + 1)]^{1/2} \\
&\qquad \qquad \times
\begin{Bmatrix}
    J' & F' & I \\
    F & J & 1
\end{Bmatrix}
(-1)^{J - \Omega}
\begin{pmatrix}
    J & 1 & J' \\
    -\Omega & q & \Omega'
\end{pmatrix}
[(2J' + 1)(2J + 1)]^{1/2}, \\
&\qquad \qquad \textrm{where }p\textrm{ describes the field polarization}\\[10 pt]
\end{aligned}
\end{equation}
\begin{equation}
\begin{aligned}
&\langle v,~ \Omega, J, I, F, M_F|\mathcal{H}_{Q}|v, \Omega, J', I, F', M_F\rangle \\
&\qquad \qquad = ~T^2_0(\bm{\nabla E}) (-1)^{F - M_F}
\begin{pmatrix}
    F & 2 & F' \\
    -M_F & 0 & M_F
\end{pmatrix}
(-1)^{F' + J + 2 + I}[(2F' + 1)(2F + 1)]^{1/2} \\
& \qquad \qquad \times
\begin{Bmatrix}
    J & F & I \\
    F' & J' & 2
\end{Bmatrix}
(-1)^{J - \Omega}
\begin{pmatrix}
    J & 2 & J' \\
    -\Omega & 0 & \Omega
\end{pmatrix}
[(2J' + 1)(2J + 1)]^{1/2} \\
& \qquad \qquad \times \langle v, \Omega|T^2_0(\bm{Q})|v, \Omega\rangle
\end{aligned}
\end{equation}

Without experimental data for TeH\+, we are forced to estimate some of the interaction constants (Table~\ref{table:constants}). In the case of hydrides, $c_I$ can be somewhat reliably predicted for the heavy atom's nuclear spin-rotation interaction (Eq. 8-41 in~\cite{townes2013microwave} or~\cite{Okabayashi2006}); however, the proton nuclear spin-rotation interaction is both difficult to observe and difficult to predict. We instead estimate the value based on measurements made for molecules possessing a heavy atom both one row below and above tellurium in the periodic table. For ZnH, $c_I(H)$ was measured to be 60 kHz~\cite{Tezcan1997}, and for AuH it was not observed within the experimental uncertainty of 30 kHz~\cite{Okabayashi2006}. A measurement with similar uncertainty was made for AsH, where the value of $c_I(H)$ was determined to be smaller than the uncertainty as well~\cite{Fujiwara1997}. Although the uncertainty is large on our estimate, its effect on the hyperfine splitting is small compared to the other hyperfine parameters. The hyperfine constants $b$ and $c$ were estimated from the AsH molecule~\cite{Fujiwara1997}, which has very similar electronic structure to TeH\+, with As one row above Te in the periodic table. Using arguments that the Fermi Contact parameter $b_F$ scales linearly with bond length~\cite{Ashworth1990} and that the dipolar constant $c$ scales as the inverse cube of the bond length~\cite{Brown1984}, $b$ and $c$ for TeH\+ were determined from the AsH values of -53 MHz and 13 MHz, respectively. The ratio of ground state bond lengths from TeH\+ to AsH is 1.07. The rotational g-factor $g_J$ was estimated from a measurement of SbH~\cite{Stackmann1993}, which has both a very similar reduced mass and electronic structure to that of TeH\+. Its small value indicates that the rotational Zeeman interaction will be the smallest Zeeman interaction.
\begin{table}
\caption{Constants used in matrix element calculations.}
\label{table:constants}
\begin{tabular}{lr}
Constant              &  Value\\
\hline
\hline
$c_I$            &      $\sim 10$ kHz\\
$b$            &       -50 MHz \\
$c$            &       10 MHz\\
$g_I$           &        5.58 \\
$g_J$		&	     -0.001 \\
$g_s$		&	     2\\
$\mu_N$		&	     $7.62 \times 10^{-4}$ MHz/G \\
$\mu_B$		&	     $1.40 \times 10^{-4}$ MHz/G \\
\end{tabular}
\end{table}

Proper definite-parity eigenstates were used for $\Omega$ doublets in the X$_21$ manifold.  For instance, the parity eigenstate in $X_21$ coupling to the negative parity $|v=0, J_a=0; \Omega = 0, J=1\rangle$  state will be
\begin{equation} \label{parity}
|v=0, J_a=1; J=1, -\rangle = \frac{1}{\sqrt{2}}(|v=0, J_a=1; \Omega = 1, J=1\rangle - |v=0, J_a=1; \Omega = -1, J=1\rangle).
\end{equation}
We also verified that including Stark couplings at expected stray field levels did not affect the Zeeman shift results.

\subsection{Quadratic Zeeman Shifts}
From estimates of the parameters above, we expect a $J=1/2$ hyperfine splitting of $\Delta = \sim 600$ kHz.  The perturbation theory expectation that the \TeH\ quadratic Zeeman shift is of order $(g_F M_F \mu_B)^2/ h \Delta$ is in good agreement with the matrix diagonalization result.  Compared with Yb$^+$, which also has field-mixed hyperfine structure, \TeH\ has a significantly smaller magnetic moment but also a much smaller hyperfine spacing.  The resulting \TeH\ quadratic Zeeman shift is similar to that of the Yb$^+$ (E2) transition and an order of magnitude larger than for the Yb$^+$ (E3) transition.

\section{Quadrupole Shifts}
The molecular quadrupole moment will cause an energy shift when coupled to a laboratory electric field gradient.  The quadrupole moment tensor T$^2_0($\textbf{Q}$)$ can be represented in Cartesian coordinates via
\begin{equation} \label{spherical}
T^2_0(\bm{Q}) = \frac{1}{\sqrt{6}}(2Q_{ZZ} - Q_{XX} - Q_{YY}).
\end{equation}
Integrating over the internuclear distance R, the quadrupole moment functions Q$_{XX}$(R), Q$_{YY}$(R) and Q$_{ZZ}$(R) for $v=8$ in X$_10^+$ yield 2.24, -1.12 and -1.12 ea$_0^2$, respectively.

\bibliography{prospects}

\end{document}